\documentclass[12pt,a4paper,final]{iopart}

\usepackage{iopams}  
\usepackage[breaklinks=true,colorlinks=true,linkcolor=blue,urlcolor=blue,citecolor=blue]{hyperref}
\usepackage[pdftex]{graphicx}
 \usepackage{tikz}
\usetikzlibrary{decorations.markings}

\begin{document}

\title[Chiral magnetic conductivity and surface states of  TI ultra-thin film multilayer]{Chiral magnetic conductivity and surface states of Weyl semimetals in topological insulator  ultra-thin film multilayer}

\author{S. A. Owerre $^{1,2}$}
\address{$^1$ African Institute for Mathematical Sciences, 6 Melrose Road, Muizenberg, Cape Town 7945, South Africa.}
\address{$^2$Perimeter Institute for Theoretical Physics, 31 Caroline St. N., Waterloo, Ontario N2L 2Y5, Canada.}
\ead{sowerre@perimeterinstitute.ca}

%

\begin{abstract}
We investigate an ultra-thin film of topological insulator (TI) multilayer as a model for a three-dimensional (3D) Weyl semimetal. We introduce tunneling parameters $t_S$, $t_\perp$, and $t_D$, where the former two parameters  couple layers of the same thin film at small and large momenta, and the latter parameter couples  neighbouring thin film layers along the $z$-direction. The Chern number is computed in each topological phase of the system and we find that for $t_S,t_D>0$, the tunneling parameter $t_\perp$ changes from positive to negative as the system transits from  Weyl semi-metallic phase to insulating phases. We further study the chiral magnetic effect (CME) of the system in the presence of a time dependent magnetic field. We compute the low-temperature dependence of the chiral magnetic conductivity and show that it captures three distinct phases of the system separated by plateaus.   Furthermore,  we propose and study a 3D lattice model of Porphyrin thin film, an organic material known to support topological Frenkel exciton edge states.  We show that this model exhibits a 3D  Weyl semi-metallic phase and also supports a 2D Weyl semi-metallic phase.   We further show that this model recovers that of 3D Weyl semimetal in topological insulator thin film  multilayer. Thus,  paving the way  for  simulating a 3D Weyl semimetal in topological insulator thin film multilayer. We obtain the surface states (Fermi arcs) in the 3D model and the chiral edge states in the 2D model and analyze their topological properties.

\end{abstract}

\vspace{2pc}

\submitto{\ J. Phys: Condens. Matter}

\section{Introduction}

Weyl semimetals are the new quantum state of matter in which a three-dimensional (3D) gapless system exhibits a nontrivial topology \cite{aab,aab0,gab, aab1, aab2}.  However,   materials that host Weyl fermions in three dimensions must break either time-reversal ($\mathcal{T}$) or inversion ($\mathcal{I}$) symmetry \cite{aab,aab0,gab}.  This guarantees that two Weyl points separated in momentum space are topologically stable \cite{vol}. They can only annihilate each other. For isolated Weyl points, the low-energy Hamiltonian  is governed by a massless Dirac equation \cite{mur,wan} $ \mathcal{H}(\bold k)=\chi v_F\boldsymbol{\sigma}\cdot(\bold{k}-\bold{k}_W^\pm)$, where $\boldsymbol{\sigma}$ is the triplet Pauli matrices and $\bold{k}$ is a 3-component Brillouin zone momentum and $\bold{k}_W^\pm$ are the locations of the Weyl points with chirality $\chi=\pm$ . Weyl points of this form are robust to external perturbation as all three Pauli matrices are used up in $\mathcal{H}(\bold k)$. The chirality of the Weyl points are related to the topological charges of the system. They act as monopole and anti-monopole of the   Berry curvature in the Brillouin zone (BZ) with  point-like Fermi arcs surface states \cite{wan}.  

 In recent years, several theoretical  proposals of Weyl semimetals have been studied. These proposals range from pyrochlore iridates \cite{wan, kre}, topological insulator (TI) multilayer  \cite{aab,aab0,gab, aab1, aab2}, magnetically doped topological band insulators \cite{liu}  to tight binding models \cite{hui5, hui6,hui7, hui8, hui9, hui10}. Recently, Weyl semimetal has been discovered experimentally in photonic crystals \cite{llu}.  The experimental realization of  Weyl semimetal in TaAs has also been reported  using angle-resolved photoemission spectroscopy \cite{Xu,lv, lv1}. In a lattice model, it is possible to generate  massless Dirac fermions with chirality in 2 dimensions. Such systems have been dubbed 2D Weyl semimetals \cite{cas}. They appear as chiral relativistic fermions  \cite{cas} and exhibit  an additional hidden discrete symmetry represented by an anti-unitary operator. The degeneracy of the resulting Weyl nodes are protected provided that there exists an anti-unitarity operator that commutes with the Hamiltonian whose square is equal to $-1$ at the degenerate points. This is reminiscent of time-reversal symmetry protected Dirac points in graphene. 

In this paper, we study  two ultra-thin film models. Firstly, we study an ultra-thin film of TI multilayer by utilizing the explicit expression of the conventional 2D TI ultra-thin film  Hamiltonian \cite{hai1,hui,hui1, hui2}, which contains quadratic corrections in its low-energy Hamiltonian, with tunneling parameters $t_\perp,t_S$. As a customary procedure, we construct a 3D version of this model by sandwiching a normal insulator between layers of TI thin film with tunneling parameter $t_D$ and a magnetic field along the $z$-direction. The resulting 3D model exhibits  topological properties similar to Burkov and Balents model \cite{aab}. However, in the present model we compute the explicit expressions for the Chern numbers in all the topological phases and show that when $t_S,t_D>0$, the tunneling parameter $t_\perp$ changes sign as the system transits from Wely semi-metallic phase to insulating phases.  We further study the low-temperature dependence of the chiral magnetic effect (CME) by computing the explicit expressions for the response function in the presence of a time-dependent magnetic field. In this case, the model does not possess any analytical solution. We numerically show that the chiral magnetic conductivity exhibits plateaus which separate three distinct phases of the system even though it is not an integer quantized quantity. 

  Secondly, we study a simple lattice model using the layers of porphyrin thin films \cite{joel}-- an organic material that can be synthesized in the laboratory. We present a detail analysis of this model in both 2 and 3 dimensions.  In particular,   we show that this   lattice model captures a 2D Weyl semi-metallic phase, whose nodes are protected by an anti-unitary operator.  In addition, our model also  captures a 3D Weyl semi-metallic phase, which appears as an intermediate phase between a 3D quantum anomalous Hall (QAH) insulator and a normal insulator (NI). It is  also shown that the porphyrin lattice model can be used as a tight binding model for topological insulator thin film multilayer. We use this model to simulate the chiral edge states of the 2D system and the surface states (Fermi arcs) of the 3D system in all the nontrivial topological phases of the system. 

\section { Topological insulator ultra-thin film multilayer}
In 2D topological insulator ultra-thin film,  the hybridization between the top and the bottom layers gives rise to a massive Dirac fermion \cite{hai1,hui,hui1, hui2}. Here, we work from this 2D low-energy Hamiltonian and construct a 3D model for Wely semimetal by  inserting insulator spacer layers between TI thin films and introduce a tunneling parameter  that couples neighbouring layers of the ultra-thin film.   The Hamiltonian for this multilayer is  given by 

\begin{equation}
H= \sum_{\bold{k}_\perp,ij}c^\dagger_{\bold{k}_\perp i}\mathcal{H}_{ij}c_{\bold{k}_\perp j},
 \end{equation}
 where
 \begin{eqnarray}
\mathcal{H}_{ij}&= v_F(\hat{z}\times \boldsymbol{\sigma})\cdot\bold{k}_\perp\delta_{ij} + (\frac{t_S}{2}-t_\perp k_\perp^2)\tau_z\sigma_z\delta_{ij} +\gamma\sigma_z\delta_{ij} \nonumber\\&+ \frac{t_D}{2}\frac{(\delta_{j,i+1}+\delta_{j,i-1})}{2}\tau_z\sigma_z.
 \label{2}
 \end{eqnarray}
  The difference between this Hamiltonian and that of  Burkov and Balents \cite{aab} is that Eq.~\ref{2} is quadratic in the momentum variables and the couplings are diagonal in the pseudo spin space. It also has an advantage in that the surface states can be simulated through a lattice model and  the Chern numbers can be obtained explicitly in all the topological phases of the system. 
  
   The Pauli matrices  $\boldsymbol{\sigma}$  denote the real spin space and $\boldsymbol{\tau}$ are the {\textit which surface} pseudo spins; $\bold{k}_\perp=(k_x,k_y)$ is a 2D momentum vector in the BZ. The indices $i,j$ label distinct thin film layers and $v_F$ is the Fermi velocity;  $t_S$  and $t_\perp$ are the tunneling parameters that couple the top and bottom surfaces of the same thin film layer for small $k_\perp$ and large $k_\perp$ respectively, and $\gamma=g\mu_B B$ is the Zeeman splitting which can be induced by magnetic doping or directly applying a magnetic field;  $t_D$ is the tunneling parameter  that couples the top and bottom surfaces of neighbouring thin film layers along the growth $z$-direction.   The parameters  $\gamma$, $t_\perp$, $t_S$, and $t_D$ depend on the thickness of the thin film, $t_\perp$ and $t_S$ have been determined both  numerically  \cite{hui,hui1,hui2} and experimentally\cite{hui4,zza1}. The new parameter $t_D$ can also be determined by growing the multilayer above. In the 2D model, the energy gap in the TI ultra-thin film can be  enhanced by using a thinner film. Thus, the thickness of the film can change the topology of the system. In the present model, a smaller thickness should also enhance the Weyl semimetallic state induced by  the interlayer coupling $t_D$ and the magnetic field. Without loss of generality we assume all the parameters to be positive $b, t_S,t_D>0$. However, as  will be shown in the subsequent sections,   $t_\perp$ can be positive or negative when moving from the Weyl semi-metallic phase to other phases of the system.
      
   It is expedient to  Fourier transform the Hamiltonian along the growth $z$-direction. We obtain 
\begin{equation}
\mathcal{H}(\bold{k})= v_F(\hat{z}\times \boldsymbol{\sigma})\cdot\bold{k}_\perp +  [\gamma+\hat{\Delta} (\bold k)]\sigma_z,
\label{fullti1}
\end{equation}
where 
\begin{equation} \hat{\Delta} (\bold k)=\big[ \frac{t_S}{2}-t_\perp k_\perp^2+ \frac{t_D}{2}\cos(k_zd)\big] \tau_z.\end{equation}
The Hamiltonian (Eq.~\ref{fullti1}) breaks $\mathcal T$-symmetry  due to  the magnetic field, but inversion symmetry is preserved $\mathcal I:~\mathcal{H}(\bold{k})\to \tau_z\sigma_z\mathcal{H}(-\bold{k}) \tau_z\sigma_z$.
   The eigenvalues of $\hat{\Delta}(\bold k)$ are $\pm\Delta(\bold k)$, where
$\Delta (\bold k) =\frac{t_S}{2}- t_\perp k_\perp^2+ \frac{t_D}{2}\cos(k_zd)$ and the corresponding eigenspinors are
\begin{equation}
\ket{u^\uparrow}={1\choose 0};\quad \ket{u^\downarrow}={0\choose1}.
\label{eig0}
\end{equation}
 Hence, the  Hamiltonian can be written as a $2\times 2$ massless Dirac equation (Weyl equations) given by
 
 \begin{equation}
\mathcal{H}_s(\bold{k})= v_F(\hat{z}\times \boldsymbol{\sigma})\cdot\bold{k}_\perp +m_s(\bold k)\sigma_z,
\label{par}
\end{equation}
with $m_s (\bold k)=\gamma + s\Delta (\bold k)$ and $s=\pm$ or $ \uparrow,\downarrow$. 
The form of  the $\hat{\Delta}$ function for this model affects the phases that emerged  when $\gamma=0$. For the present model, Eq.~\ref{par} with $\gamma=0$ describes a 3D Dirac semimetal which possesses  time-reversal and inversion symmetries. It exhibits a phase with two Dirac nodes along the $k_z$-direction when $t_S <t_D$ and an insulating phase ( 3D QSH phase) for $t_S>t_D$. In the insulating phase,  the $Z_2$ topological number is 
$(-1)^\nu= -\textrm{sign}(t_S-t_D)$, where $\nu=1$ characterizes a nontrivial phase. The semi-metallic phase and the insulating phase are separated by a saddle point at $k_x=k_y=0,~k_z=\pi/d$, with energy $\pm|t_S-t_D|/2$. 

 To obtain a nontrivial Weyl semi-metallic phase,  $\mathcal{T}$- or $\mathcal{I}$-symmetry must be broken as mentioned above. This requires that $\gamma\neq 0$.  The  corresponding energy eigenvalues of Eq.~\ref{par} are given by \begin{equation}
 \epsilon_{s\lambda} (\bold k)=\lambda\sqrt{v_F^2k_\perp^2+m_s^2 (\bold k)}=\lambda\epsilon_{s} (\bold k), 
\end{equation}
where $\lambda=\pm$ labels the conduction and the valence bands respectively, and the eigenvectors are
\begin{equation} \ket{v_\bold{k}^{\lambda s}} = \frac{1}{\sqrt{2}}\lb\sqrt{1+\lambda\frac{m_s (\bold k)}{\epsilon_{s}(\bold k)}},-i\lambda e^{i\phi_{\bold k_\perp}}\sqrt{1-\lambda\frac{m_s (\bold k)}{\epsilon_{s}(\bold k)}}\rb^T\label{eig1},\end{equation} where $e^{i\phi_{\bold k_\perp}} =(k_x+ik_y)/k_\perp$.
Hence, the eigenspinors of the complete system are $\ket{\psi_\bold{k}^{\lambda s}} =\ket{u^s}\otimes \ket{v_\bold{k}^{\lambda s}}$, where 
 \begin{equation}
 \ket{\psi_\bold{k}^{\uparrow\lambda}}=  {\ket{v_\bold{k}^{\uparrow\lambda}} \choose \bold{0}} ; \quad \ket{\psi_\bold{k}^{\downarrow\lambda}}=  {\bold{0}\choose\ket{ v_\bold{k}^{\downarrow\lambda}}}.
 \label{spino}
  \end{equation}
Two Weyl nodes are realized in the $\mathcal{H}_\downarrow(\bold k)$ block of the Dirac equation (Eq.~\ref{par}). This corresponds to  the solutions of $m_\downarrow(\bold k)=0$, where  $m_\uparrow(\bold k)$ never changes sign. The  Weyl nodes  are located at  $k_x=k_y=0$, $k_z^\pm= \pi/d \pm k_W$, where
\begin{equation}
k_W=\frac{1}{d}\arccos\lb 1-\frac{2}{t_D}\lb \gamma-\gamma_-\rb \rb,
\label{nod1}
\end{equation}
with $\gamma_\pm=(t_S\pm t_D)/2$ and  $\gamma_+ >\gamma_-$. 
 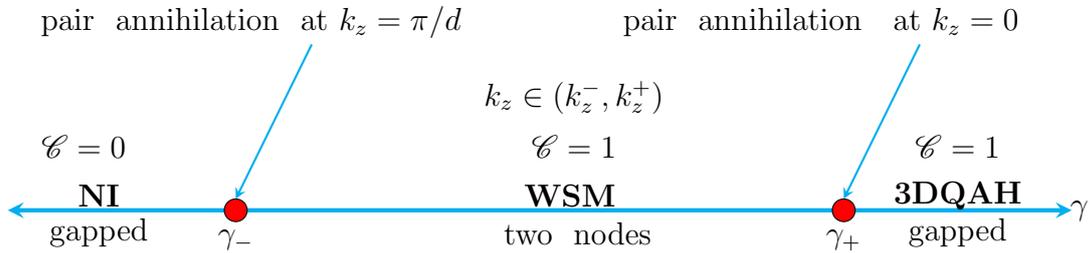
\begin{figure}[ht]
\centering
\begin{tikzpicture}
Brillouin Zone
\draw [<->,>=stealth,ultra thick, cyan] (-4,0.0) --(10,0);
\draw (10.1,0) node[]{$\gamma$};
\draw (-1,-.4) node[]{$\gamma_-$};
\draw (7,-.4) node[]{$\gamma_+$};
\draw (-2.8,.2) node[]{$\bf{NI}$};
\draw (-3,.85) node[]{$\mathscr{C}=0$};
\draw (8.5,.2) node[]{$\bf{3D QAH}$};
\draw (8.5,.85) node[]{$\mathscr{C}=1$};
\draw (3.4,.2) node[]{$\bf{WSM}$};
\draw (3.45,1.5) node[]{$k_z\in(k_z^-, k_z^+)$};
\draw (3.45,.85) node[]{$\mathscr{C}=1$};
\draw[fill= red] (-1,0) circle (1.5mm);
\draw[fill= red] (7,0) circle (1.5mm);
\draw [<-,>=stealth, thick, cyan] (7,0.15) --(8,2.2);
\draw (6.7,2.5) node[]{$\textrm{pair~ annihilation ~ at}~ k_z=0$};
\draw [<-,>=stealth, thick, cyan] (-1,0.15) --(0,2.2);
\draw (-.8,2.5) node[]{$\textrm{pair ~annihilation~ at}~k_z=\pi/d$};
\draw (-2.8,-.3) node[]{$\textrm{gapped}$};
\draw (8.5,-.3) node[]{$ \textrm{gapped}$};
\draw (3.5,-.3) node[]{$ \textrm{two~ nodes}$};
\end{tikzpicture}
\caption{Color online.  The phase diagram of the TI multilayer comprising a normal insulator (NI), Weyl semimetal (WSM), and quantum anomalous Hall (QAH) insulator  along the magnetic field line. The values of the corresponding Chern numbers are indicated by $\mathscr C$.}
\label{pha}
\end{figure}

The phase diagram in Fig.~\ref{pha}   comprises an ordinary insulator phase for $\gamma<\gamma_-$,  and   a 3D QAH phase for  $\gamma>\gamma_+$. A 3D Weyl semimetal  with two Weyl nodes appears in the regime $\gamma_-<\gamma<\gamma_+$, and a pair annihilation occurs exactly at the boundaries. As in all theoretical  models, a 3D Weyl semimetal phase always appears as an intermediate state between an ordinary insulator and a 3D quantum anomalous Hall insulator.  The Hall conductivity  is given by
\begin{equation}
\sigma_{xy}^s(k_z)=\frac{e^2}{h}\mathscr{C}^{s}(k_z).
\label{qahc}
\end{equation}
 In the present model, we can calculate the Chern number explicitly by treating $k_z$ as a parameter, thus reducing the problem to an effective 2D model. Hence, the Chern number is computed with the same formula  \cite{fuk} \begin{equation}
 \mathscr{C}^{s}(k_z)=\int d  k_\perp \frac{\partial\mathcal{A}_{\phi_{\bold{k_\perp}}}^s}{\partial k_\perp},
\label{thou3}
\end{equation}
where \begin{equation}
\mathcal{A}_{\phi_{\bold{k_\perp}}}^s=i\braket{\psi^{s}_-|\partial_{\phi_{\bold{k}_\perp}}\psi^{s}_-}.
\label{dee}
 \end{equation}
The $s=\downarrow (-)$ block  realizes  Weyl nodes, therefore the Chern number is defined only for the occupied band  of this block. Using Eqs.~\ref{thou3} and \ref{dee} we obtain 

 \begin{equation}
\mathscr{C}^{\downarrow}(k_z)=-\frac{1}{2}[\textrm{sign} \lb \tilde{m}_\downarrow(k_z)\rb- \textrm{sign}(t_\perp)],
 \label{chern}
\end{equation}
where ${\tilde{m}_\downarrow} (k_z)=\gamma-\lb\frac{t_S}{2}+ \frac{t_D}{2}\cos(k_zd)\rb.$

In the Weyl semi-metallic phase $\gamma_-<\gamma<\gamma_+$,  $k_z$ must take  values in-between the nodes, i.e., $k_z\in ( {k}^-_z,  k^+_z)$,  hence ${\tilde{m}_\downarrow} (k_z)<0$. A nonzero Chern number then requires $t_\perp>0$.  The Chern number only changes when the gap closes and reopens at the boundaries $\gamma=\gamma_\pm$. Once the gap closes and reopens  we must have $t_\perp<0$ to get a normal insulator phase and a nontrivial 3D QAH phase at $k_z=\pi/d$ and $k_z=0$ respectively (see Fig.~\ref{pha}). This can be explicitly shown by evaluating the Chern number  at $k_z=0$ where the gap closes and reopens for $\gamma>\gamma_+$.  We obtain  \begin{equation}
\mathscr{C}^{\downarrow}(0)= -\frac{1}{2}[\textrm{sign}\lb \gamma-\gamma_+\rb-\textrm{sign}\lb t_\perp\rb].
\label{QAH}
\end{equation}
A similar situation occurs at $k_z=\pi/d$ for $\gamma<\gamma_-$, and the  Chern number is given by\begin{equation}
\mathscr{C}^{\downarrow}\lb\frac{\pi}{d}\rb= -\frac{1}{2}[\textrm{sign}\lb \gamma-\gamma_-\rb-\textrm{sign}(t_\perp)].
\label{NI}
\end{equation}
 Note that  Eqs.~\ref{QAH} and \ref{NI} reduce to the 2D Chern number \cite{hai1} when $\gamma=t_D=0$. In this case, the band inversion requires that $t_\perp t_S>0$ for a nontrivial topological phase to exist. For $\gamma\neq 0$, the present model  requires that $t_\perp<0$ as mentioned above. This guarantees that the first Chern number, Eq.~\ref{QAH}, is integer quantized and describes a 3D QAH phase and the second Chern number, Eq.~\ref{NI}, is zero which describes a normal insulator phase. As mentioned above, the topological property of Weyl semimetal is also manifested as monopoles and anti-monopoles of the Berry curvature. This is evident by expanding Eq.~\ref{chern} near the Weyl nodes, we obtain
\begin{equation}
\mathscr{C}^{\downarrow}(k_z)=\chi\tilde{v}_F\textrm{sign}(k_z-k_z^\pm)+\cdots
\end{equation}
where $\tilde{v}_F=t_D d\sin(k_Wd)/4$, and $\chi=\pm$ is the chirality of the Weyl nodes.  This expression explicitly shows a monopole and anti-monopole at $k_z=k_z^\pm$ with chirality $\chi=\pm $ respectively.

The Fermi arcs in the vicinity of the Weyl nodes is a special feature of Weyl semimetals. These arcs are exactly the edge states of the effective 2D system for fixed $k_z$, and exist for any surface not perpendicular to the $z$-axis \cite{aab}.  We can  explicitly  solve for these edge states by considering a slab geometry   occupying the half-plane $x\geq0$ with open boundary condition along $x$-direction and translational invariant in the $y$-$z$ plane. Thus, $k_y$ and $k_z$ are good quantum numbers and $k_x$ is replaced by $k_x\to-i\partial_x$. The Hamiltonian can be written as
\begin{equation}
H_-=i\sigma_y v_F\partial_x-\sigma_zt_\perp\partial_x^2 +v_F\sigma_x k_y -\sigma_z t_\perp k_y^2\nonumber\\ +m_-(k_z,x)\sigma_z.
\label{bulk}
\end{equation}
We first consider $k_y=0$ and solve for the zero energy solution of the Schr\"{o}dinger equation $\mathcal{H}_-\Phi(k_z,x)=0$,
\begin{equation}
[ v_F\partial_x +t_\perp\sigma_x\partial_x^2-m_-(k_z,y)\sigma_x]\Phi(k_z,x)=0,
\label{xx1}
\end{equation}
where  $\Phi(k_z,x)$ is a 2-component spinor and we have multiplied through by $-i\sigma_y$. We seek for a solution of the form
\begin{equation}
\Phi_\lambda(k_z,x)=\xi_\lambda e^{\omega x},
\end{equation}
where  $\sigma_x\xi_\lambda=\lambda\xi_\lambda$, ($\lambda=\pm1$) and $\omega$ solves the equation
\begin{equation}
v_F\omega+ \lambda t_\perp\omega^2- \lambda m_-(k_z,x)=0.
\label{ome}
\end{equation}
The allowed solution that obey the boundary conditions of the wavefunction [$\Phi(k_z,0)=\Phi(k_z,\infty)=0$] is given by
\begin{equation}
\Phi_\lambda(k_z,x)= \frac{\mathcal C}{\sqrt{2}}{1 \choose \lambda}(e^{-\omega_+ x}-e^{-\omega_- x}),
\end{equation}
 where $\mathcal C$ is a normalization constant, and $\omega_\pm$ are the positive solutions of Eq.~\ref{ome}. The surface Hamiltonian is obtained by projecting Eq.~\ref{bulk} onto the surface states
\begin{equation}
\mathcal{H}_{sur}(k_y,k_z)=\Phi_\lambda^\dagger \mathcal{H}_- \Phi_\lambda= v_Fk_y\sigma_z.
\end{equation}

\section {Magnetic field response}
In the previous section, we derived the phase diagram, anomalous Hall conductivity,  and surface states of an ultra-thin film of TI Hamiltonian with quadratic momentum corrections. In this section, we study the response of the system to an orbital magnetic field through the  vector potential, $\bold{A}=x\mu_B B\hat{y}$, which corresponds to a magnetic field along the growth $z$-direction. The Hamiltonian is given by
  \begin{equation}
\mathcal{H}_s(\bold{k})= v_F(\hat{z}\times \boldsymbol{\sigma})\cdot\lb -i\boldsymbol{\nabla} +e\bold{A}\rb+m_s(\bold k)\sigma_z.
\label{par1}
\end{equation}
We introduce the operator $\boldsymbol{\pi}=-i\boldsymbol{\nabla} +e\bold{A}$, and define the creation and annihilation operators:
$
 a=l_B(\pi_x-i\pi_y)/\sqrt{2}; ~ a^\dagger=l_B(\pi_x+i\pi_y)/\sqrt{2},
$
where $l_B^2=(e\mu_BB)^{-1}$ is the magnetic length. 
In terms of $a$ and $a^\dagger$ the Hamiltonian becomes
 \begin{equation}
\mathcal{H}_s(k_z)=i\omega_B\sqrt{2}(\sigma^+a-\sigma^-a^{\dagger})+m_s(k_z)\sigma_z,
\label{hamm}
\end{equation}
 where $\sigma^\pm =(\sigma_x\pm i\sigma_y)/2$ and $m_s(k_z)=\gamma+s{\Delta}(k_z)$, where ${\Delta}(k_z)$ is given by \begin{equation}
{\Delta}(k_z)=\bigg[\frac{t_S}{2}+ \frac{t_D}{2}\cos(k_zd)-\omega_0\lb a^\dagger a+ \frac{1}{2}\rb\bigg].
\end{equation}
Here, $\omega_B={v_F}/l_B$ is the magnetic frequency and $\omega_0= 2t_\perp/{l_B^2}$ is the harmonic oscillator frequency. The eigenvector of each $2\times 2$ block may be written as
    
  \begin{equation}
\mathcal{U}_{sn}={\alpha_{sn}^1u_{n-1}\choose \alpha_{sn}^2u_{n}},
\end{equation}
where $\alpha_{sn}^1,\alpha_{sn}^2$ are constants to be determined. The operators satisfy $au_{n}=\sqrt{n}u_{n-1}$; $a^\dagger u_{n}=\sqrt{n+1}u_{n}$. Hence, Eq.~\ref{hamm} yields a $2\times 2$ eigenvalue equation for $\alpha_{sn}$. The Hamiltonian yields
 \begin{equation}
\mathcal{H}_{sn}(k_z)=s\frac{\omega_0}{2}\sigma_0 -\omega_B\sqrt{2n}\sigma_y+m_{sn}(k_z)\sigma_z,
\label{hamm1}
\end{equation}
where
$\sigma_0$ is an identity matrix, $m_{sn}(k_z)=\gamma+s{\Delta}_n(k_z)$, and \begin{equation}
{\Delta}_n(k_z)=\frac{t_S}{2}+ \frac{t_D}{2}\cos(k_zd)-\omega_0 n.
\end{equation}
 The eigenvalues of Eq.~\ref{hamm1} are given by

\begin{figure}[ht]
\centering
\includegraphics[width=4in]{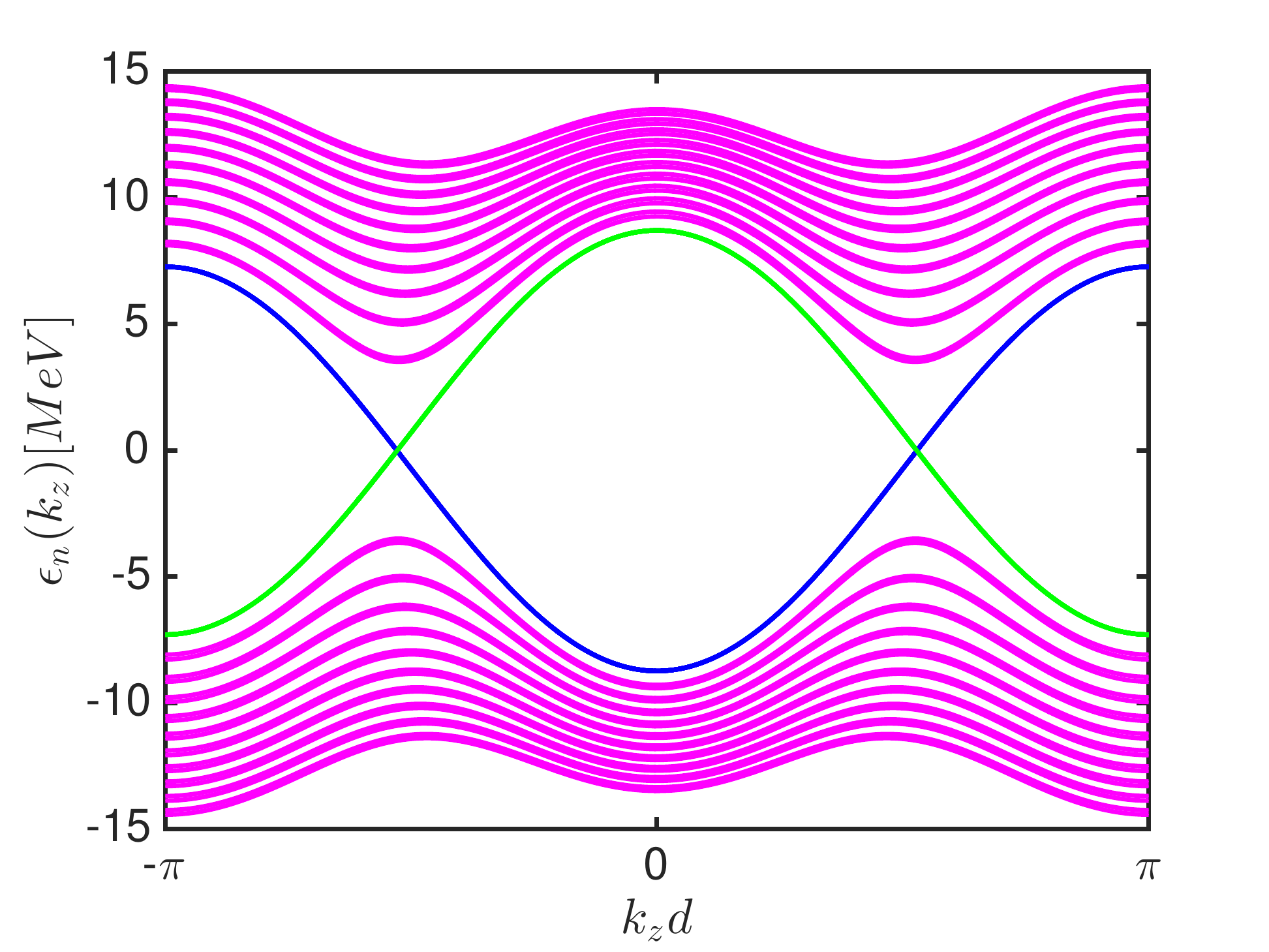}
\caption{Color online. The Landau levels as a function of the momentum $k_zd$. The parameters are $\mu_BB=0.1$T; $t_S= 1$MeV, $t_D= 10$MeV, $v_F=500$MeV; $g=0.1$MeV.}
\label{LL}
\end{figure}

 \begin{eqnarray}
&\epsilon_{n,\lambda}^s (k_z)=s\frac{\omega_0}{2}+\lambda\sqrt{2n\omega_B^2 +m_{sn}^2(k_z)},\quad n\geq 1,\label{eqnn}\\&
 \epsilon_{0}^s(k_z)=s\frac{\omega_0}{2}-m_{s,0}(k_z), \quad n=0.
 \end{eqnarray}
  The corresponding eigenvectors are
\begin{eqnarray} 
&\ket{v_{n}^{\lambda s}} = \frac{1}{\sqrt{2}}\lb \sqrt{1+\lambda\frac{m_{sn} (k_z)}{\tilde{\epsilon}_{n}^s (k_z)}},-i\lambda\sqrt{1-\lambda\frac{m_{sn} (k_z)}{\tilde{\epsilon}_{n}^s (k_z)}}\rb^T,\label{eig1}\\& \ket{v_{0}^s}=(0,1)^T,\end{eqnarray} where $\tilde{\epsilon}_{n\lambda}^s (k_z)=\epsilon_{n\lambda}^s (k_z)+s\omega_0/2$. The eigenspinors of the complete system are $\ket{\psi^{s\lambda}}=\ket{u^s}\otimes \ket{v_{n}^{s\lambda}}$, where 
 \begin{equation}
 \ket{\psi_{n}^{\uparrow \lambda}}=  {\ket{v_{n}^{\uparrow \lambda}} \choose \bold{0}} \quad\textit{and} \quad \ket{\psi_{n}^{\downarrow\lambda}}=  {\bold{0}\choose \ket{v_{n}^{\downarrow\lambda}}}.
 \label{spino}
  \end{equation} 
For $\gamma=0$, the zero Landau levels crosses at $ B_c= (t_S+t_D\cos(k_zd))/2e\mu_Bt_\perp$, which vanishes at the Dirac nodes $k_z^\pm=\pi/d\pm k_W$. At the transition point $k_z=\pi/d$ ,  $B_c\neq 0$. For $t_S>t_D$, the regime $B<B_c$ corresponds to a 3D QSH phase and $B>B_c$ corresponds to trivial phase.
  The Landau level for $\gamma\neq 0$ is shown in Fig.~\ref{LL}, which evidently captures the appearance of two Weyl nodes in the vicinity of the bulk gap.
 \subsection{Chiral magnetic effect}
 Chiral magnetic effect is the response of a system to a  time-dependent magnetic field. This phenomenon is well-known  in high energy physics as  the chiral magnetic conductivity. For instance, gluon field configurations with nonzero topological charges exhibit this effect \cite{dm}. It has been shown to occur in  Weyl semimetals \cite{aab2,bur}.   In this subsection, we  investigate the low-temperature dependence of the chiral magnetic conductivity on the TI ultra-thin film  Hamiltonian. We will derive the expressions for our model, which do not possess any analytical solution. We also show that the  chiral magnetic conductivity captures the appearance of the three distinct phases of the system though it is not integer quantized like the quantum anomalous Hall conductivity.   In the linear response theory, the current operator is given by
 \begin{equation}
 J^i(\bold{q}, \omega)= \Pi_{ij}(\bold{q}, \omega) A^j(\bold{q}, \omega),
 \end{equation}
 where $\Pi_{ij}(\bold{q}, \omega)$ is the current-current correlation function.
The chiral magnetic effect (or conductivity) arises in the presence of a time-dependent magnetic field along the $z$-direction. In the Landau gauge we adopt here, the magnetic field is only related to the $A_y$ component of the gauge field, that is  $B_z=\partial_x A_y(x)$. Assuming $A_y(x)= A(\bold q, \omega) e^{i(qx-\omega t)}$, we have $B_z(\bold q,\omega)=iq A(\bold q, \omega)$. The response of the system to a time-dependent magnetic field gives rise to an induced current given by
\begin{equation}
J(\bold q, \omega)= \sigma_{\chi}(\bold q, \omega) B_z(\bold q, \omega).
\end{equation}
Thus, the chiral magnetic   conductivity is 
\begin{equation}
 \sigma_{\chi}(\bold q, \omega)=\frac{-i}{q}\Pi( \bold q, \omega).
 \label{condd}
\end{equation}
The response function $\Pi(\bold{q}, \omega)$ is in general antisymmetric. The most convenient way to calculate the response function is from the imaginary time path integral of Eq.~\ref{par1} minimally coupled to a vector potential, 
\begin{equation}
\mathcal{S}=\int d\tau d^3 r~\psi^\dagger(\bold r, \tau)\Big[ \partial_\tau -\mu +ie A_0(\bold r, \tau)+\mathcal{H}_s\Big] \psi(\bold r, \tau).
\end{equation}
After integrating out the fermion degree of freedom, the response function is given by \cite{bur}
 \begin{eqnarray}
 -i\Pi(\bold q, i{\Omega})&=i\frac{e^2v_F}{V}\sum_{ss^\prime;\lambda\lambda^\prime}\sum_{\bold k}\frac{f[\xi_{s^\prime\lambda^\prime}(\bold{k})]-f[\xi_{s\lambda}(\bold{k+q})]}{i\Omega +\xi_{s^\prime\lambda^\prime}(\bold{k})-\xi_{s\lambda}(\bold{k+q})}\nonumber\\&\times\braket{\psi^{s\lambda}_{\bold{k+q}}|\psi^{s^\prime\lambda^\prime}_{\bold{k}}}\braket{\psi^{s^\prime\lambda^\prime}_{\bold{k}}|\boldsymbol{\sigma}\cdot\bold{q}|\psi^{s\lambda}_{\bold{k+q}}},
 \label{res}
 \end{eqnarray}
where $i\Omega=\omega+i\eta$ and $f[\xi_{s\lambda}(\bold{k})]=[e^{\xi_{s\lambda}(\bold{k})/T}+1]^{-1}$ is the Fermi function, with $\xi_{s\lambda}(\bold{k})=\epsilon_{s\lambda}(\bold{k})-\epsilon_F$. Without loss of generality we assume $\epsilon_F>0$. The spatial contribution only comes from the Landau gauge choice, thus we take $\bold{q}=q\hat{x}$. There are two contributions to the response function --- the interband with $\lambda\neq \lambda^\prime$ and the intraband with $\lambda= \lambda^\prime$. We are interested in the low-frequency and long wavelength limits $i\Omega\to 0 (\textit{second limit});~q\to 0 (\textit{first limit})$ and $q\to 0  (\textit{second limit});~ i\Omega\to 0  (\textit{first limit})$.  However, the two limits are not commutative so the order in which the limits are taken is very crucial. The former limit is the direct current (DC) limit of a transport coefficient, while the latter is the static limit. For the interband case, both order of limits contribute to the response function, so we can start with $i\Omega=0$. In this case all other terms in Eq.~\ref{res} are finite as $q=0$ except $\braket{\psi^{s\lambda}_{\bold{k+q}}|\psi^{s^\prime\lambda^\prime}_{\bold{k}}}$. Hence, we will expand this term to first order in $q$. Since the pseudo spin scalar product produces a term $\braket{u^s|u^{s^\prime}} =\delta_{ss^\prime}$, we have
\begin{eqnarray}
\braket{\psi^{s\pm}_{\bold{k+q}}|\psi^{s^\prime\mp}_{\bold{k}}}&=\delta_{ss^\prime}\frac{q}{2v_Fk_\perp\epsilon_s^2(\bold k)}\Big[\pm 2k_xt_\perp m_s^2(\bold k)\nonumber\\&+ v_F^2\lb -ik_y\epsilon_s(\bold k)\mp k_x m_s(\bold k)\rb],\label{res1}\\\braket{\psi^{s^\prime\pm}_{\bold{k}}|\psi^{s\mp}_{\bold{k}}}&=\delta_{ss^\prime}\frac{1}{k_\perp}[-k_y\frac{\epsilon_s(\bold k)}{ m_s(\bold k)}\pm i k_x].
\label{res2}
\end{eqnarray}
Plugging Eqs.~\ref{res1} and \ref{res2} into Eq.~\ref{res}, the terms containing $k_xk_y$ vanish by angular integration, we obtain 
 \begin{equation}
 -i\Pi^{inter}(\bold q, i{\Omega})=\frac{e^2q}{2}\sum_s \int \frac{d\bold k}{(2\pi)^3}\frac{1-f[\epsilon_{s}(\bold{k})-\epsilon_F]}{\epsilon_{s}^3(\bold{k})}M_s(\bold k)\label{res3},
 \end{equation}
where
\begin{equation}
M_s(\bold k)= v_F^2m_s(\bold k)-2\frac{k_x^2}{k_\perp^2} t_\perp m_s^2(\bold k).
\end{equation}
Performing the angular integration, we obtain
\begin{eqnarray}
 &-i\Pi^{inter}(\bold q, i{\Omega}) =-\frac{e^2q}{8\pi^2}\sum_s\int_{-\pi/d}^{\pi/d} dk_z\int_{0}^\infty dx\label{res4} [\Omega_{1z}^s(x,k_z)+\Omega_{2z}^s(x,k_z)]\nonumber\\&\times[1-f(\sqrt{x+m_s^2(x,k_z)}-\epsilon_F)],
\end{eqnarray}
where $x=v_F^2k_\perp^2$ and
\begin{eqnarray}
\Omega_{1z}^s(x,k_z)&= -\frac{m_s(x,k_z)}{2[x +m_s^2(x,k_z)]^{3/2}},\nonumber\\ \Omega_{2z}^s(x,k_z)&= \frac{\tilde{t}_\perp m_s^2(x,k_z)}{2[x +m_s^2(x,k_z)]^{3/2}},\\\nonumber m_s(x,k_z)&=\gamma+s[\frac{t_S}{2}-\tilde{t}_\perp x +\frac{t_D}{2}\cos(k_zd)],
\end{eqnarray}
with $\tilde{t}_\perp= t_\perp/v_F^2$.   
Now for the intraband case $\lambda=\lambda^\prime$, the response function vanishes in the  DC limit $i\Omega\to 0;~q\to 0$, i.e., if we take long-wavelength limit first. However, in the static limit $q\to 0,~i\Omega\to 0$, it is nonzero. In this case, we have
\begin{eqnarray}
&f[\xi_{s\lambda}(\bold{k+q})]=f[\xi_{s\lambda}(\bold{k})] +\bold{q}\frac{\partial \xi_{s\lambda}(\bold{k}) }{\partial \bold{q}} \frac{\partial f[\xi_{s\lambda}(\bold{k})] }{\partial \xi_{s\lambda}(\bold{k})}+\cdots\\&\xi_{s\lambda}(\bold{k+q})=\xi_{s\lambda}(\bold{k})+\bold{q}\frac{\partial \xi_{s\lambda}(\bold{k}) }{\partial \bold{q}}+\cdots
\end{eqnarray}

The intraband response function is given by

\begin{equation}
-i\Pi^{intra}(\bold q, i{\Omega})=\frac{e^2q}{2}\sum_s\int \frac{d\bold k}{(2\pi)^3}\lb -\frac{\partial f[\xi_{s\lambda}(\bold{k})] }{\partial \xi_{s\lambda}(\bold{k})}\rb\frac{M_s(\bold k)}{\epsilon^2_s(\bold k)}.
\label{intra}
\end{equation}
 In the present model, the integrations [Eqs.~\ref{res4} and \ref{intra}] cannot be done analytically. 
 We can reduce the problem in a way that is amenable to numerical integration by performing the angular integration. We obtain
\begin{eqnarray}
 -i\Pi^{intra}(\bold q, i{\Omega})& =-\frac{e^2q}{8\pi^2}\sum_s\int_{-\pi/d}^{\pi/d} dk_z\int_{0}^\infty dx[\tilde{\Omega}_{1z}^s(x,k_z)+\tilde{\Omega}_{2z}^s(x,k_z)]\label{res5}\nonumber\\& \times\Big[4T\cosh^2\lb\sqrt{x+m_s^2(x,k_z)}-\epsilon_F\rb\Big]^{-1},
\end{eqnarray}
where

\begin{equation}
\tilde{\Omega}_{1z}^s(x,k_z)= -\frac{m_s(x,k_z)}{2[x +m_s^2(x,k_z)]};~ \tilde{\Omega}_{2z}^s(x,k_z)= \frac{\tilde{t}_\perp m_s^2(x,k_z)}{2[x +m_s^2(x,k_z)]}.
\end{equation}
The conductivity is given by Eq.~\ref{condd}. In the two non-commutative limits we obtain two conductivities given by
\begin{eqnarray}
\sigma_{\chi}&= \lim_{{\Omega}\to 0}\lim_{q\to 0}\frac{-i}{q}\Pi^{inter}(\bold q, i{\Omega})\label{cme},\\\sigma_{t}&= \lim_{q\to 0}\lim_{{\Omega}\to 0}\frac{-i}{q}[\Pi^{inter}(\bold q, i{\Omega})+\Pi^{intra}(\bold q, i{\Omega})].
\label{th}
\end{eqnarray}
The first limit [Eq.~\ref{cme}] is the chiral magnetic effect (CME).  As mentioned above, this is nothing but  the chiral magnetic conductivity, a phenomenon well-studied in high energy physics \cite{dm}.  The second limit [Eq.~\ref{th}]  is a thermodynamic equilibrium quantity corresponding to the static limit;  we will focus on Eq.~\ref{cme}.
 \begin{figure}[ht]
\centering
\includegraphics[width=4in]{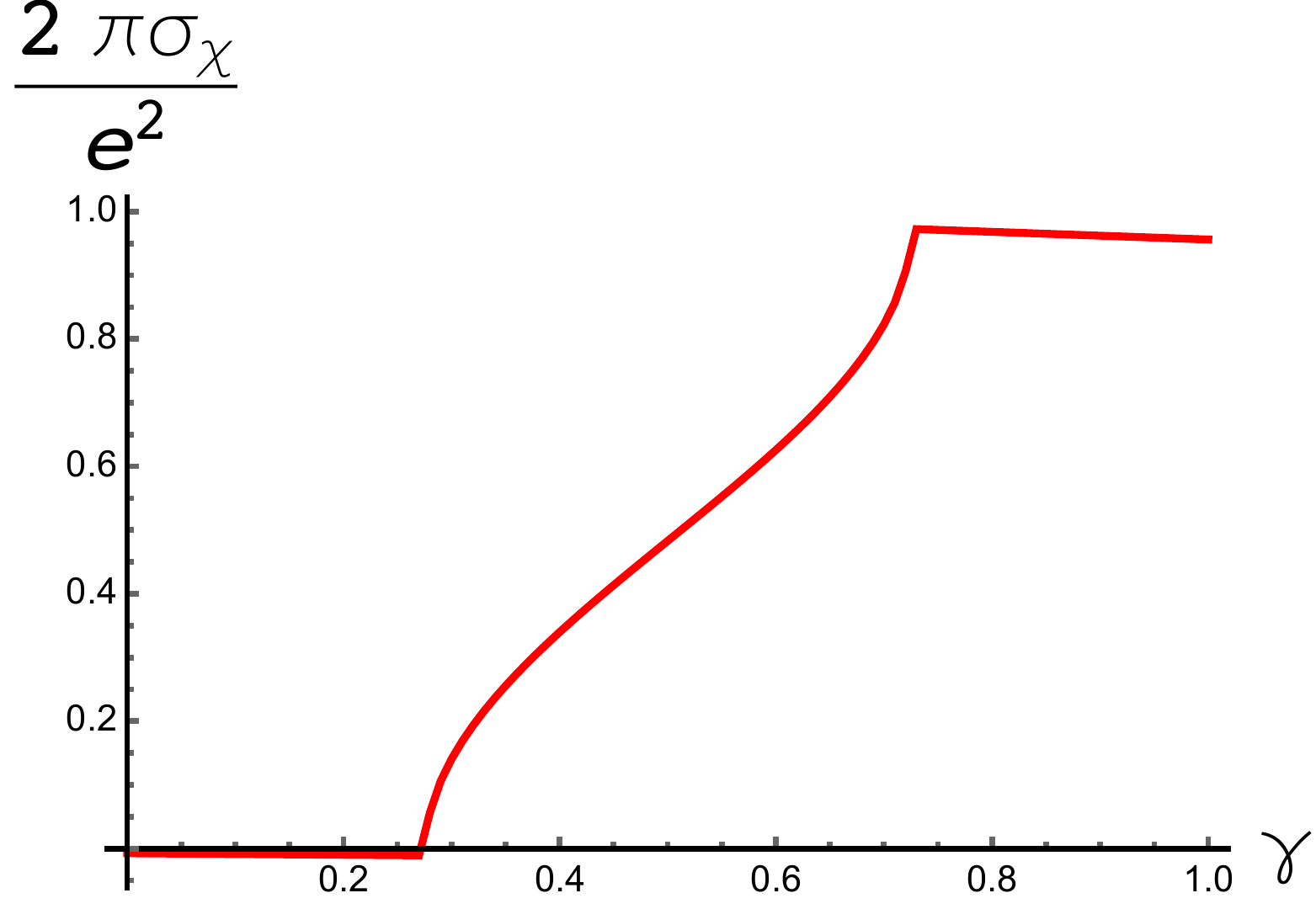}
\caption{Color online. The chiral magnetic conductivity as a function of  $\gamma$ at $\epsilon_F=0$ and $T=0.001$ in units of $t_S$. The parameters are $t_D=0.45t_S$ and $t_\perp=0.01t_S$. The sign of $t_\perp$ is unimportant because $m_\pm$ contribute to $\sigma_\chi$. The different regimes separated by plateaus are ordinary insulator $\gamma<\gamma_-$ ($\sigma_\chi=0$) ;  Weyl semimetal $\gamma_-<\gamma<\gamma_+$  ($\sigma_\chi\neq 0$), and  quantum anomalous Hall insulator $\gamma>\gamma_+$ ($\sigma_\chi\neq 0$). }
\label{cond1}
\end{figure}

\begin{figure}[ht]
\centering
\includegraphics[width=4in]{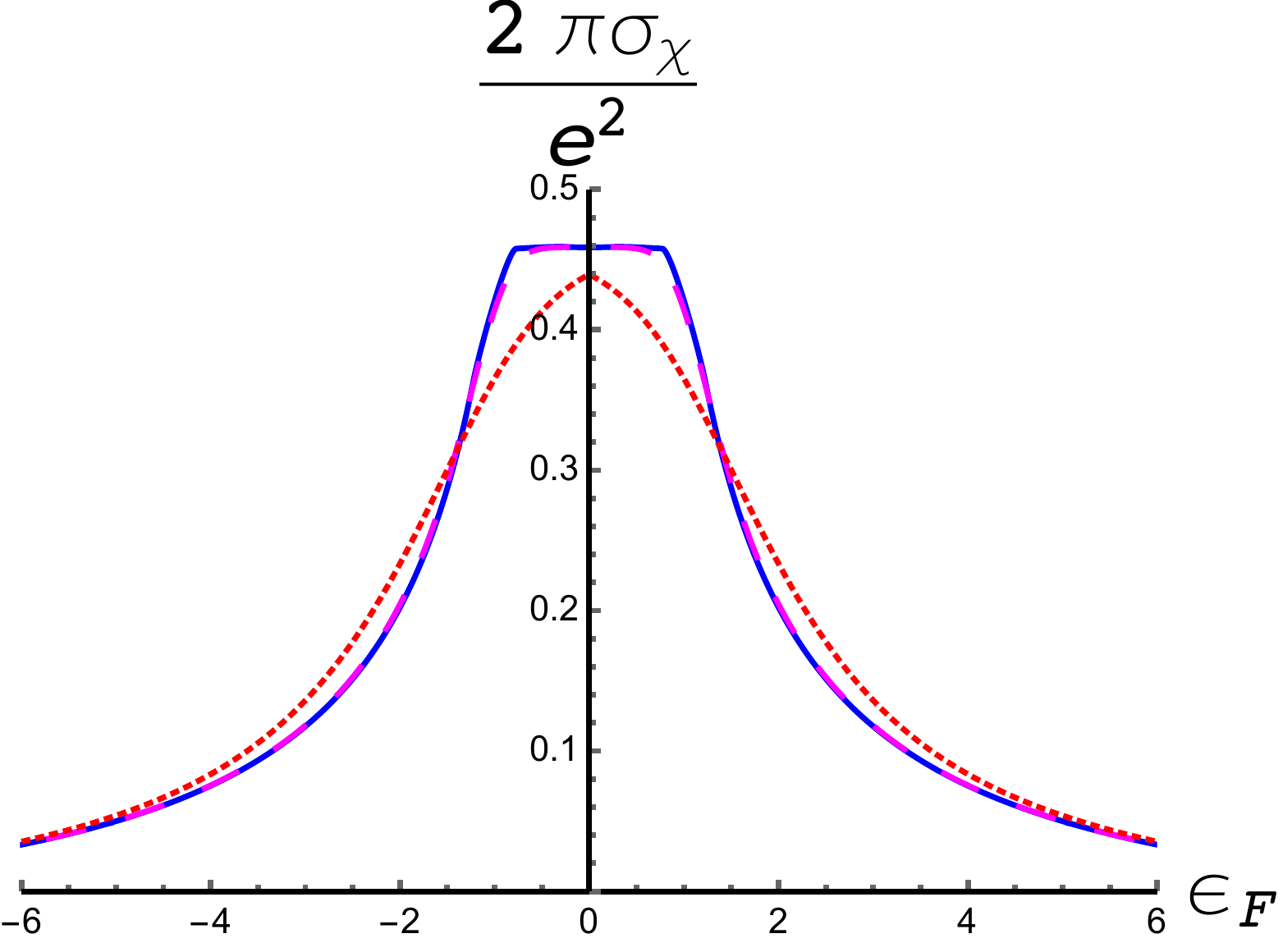}
\caption{Color online. The chiral magnetic conductivity as a function of the Fermi energy for several temperatures  $T=0.001$ (solid); $T=0.1$ (dashed); $T=0.5$ (dotted). The parameters are in units of $t_S$ with  $t_D= 0.45t_S$;  $t_\perp= 0.01t_S$, $v_F=1$;  $\gamma_-<\gamma=0.5t_S<\gamma_+$.}
\label{cond}
\end{figure}

 Figure~\ref{cond1} shows the plot of the chiral magnetic  conductivity against the magnetic field $\gamma$. Note that the sign of $t_\perp$ is irrelevant because the two masses $m_\pm(\bold k)$ contribute in the computation of chiral magnetic conductivity. Interestingly, the chiral magnetic conductivity captures the appearance of the three phases of the system. The plateaus of $\sigma_\chi$ correspond to  phase transitions from ordinary insulator $\gamma<\gamma_-$ ($\sigma_\chi=0$) to Weyl semimetal $\gamma_-<\gamma<\gamma_+$  ($\sigma_\chi\neq 0$), and from Weyl semimetal to quantum anomalous Hall insulator $\gamma>\gamma_+$ ($\sigma_\chi\neq 0$). Also notice from  Fig.~\ref{cond1} that  the chiral magnetic  conductivity is not a quantized quantity unlike the quantum anomalous Hall conductivity, Eq.~\ref{qahc}.   Figure \ref{cond} shows the chiral magnetic conductivity as a function of the Fermi energy. In this case, a step peak occurs at $\epsilon_F=0$ at low temperatures. 

\section{Porphyrin thin film multilayer}
In this section, we propose and analyze a lattice model for Weyl semimetals from a porphyrin thin film layer. We will also show the connection of this model to that of TI thin film studied above. To construct a 3D lattice model, it is customary to  stack  layers of porphyrin thin film on top of each other along the $z$-direction. The  2D Hamiltonian of a porphyrin thin film  is given by \cite{joel}

\begin{eqnarray}
&H_{2D}=\sum_{m,\alpha} \bigg[{\mathcal{J}_{1}}\lb e^{i{\Phi}} a^\dagger_{m}b_{m+\hat{\boldsymbol{\delta}}_\alpha}+e^{-i{\Phi}} a^\dagger_{m}b_{m-\hat{\boldsymbol{\delta}}_\alpha}\rb \nonumber\\& +{\mathcal{J}_{2}}\lb e^{-i{\Phi}} a^\dagger_{m}b_{m+\hat{\boldsymbol{\delta}}_\alpha}+e^{i{\Phi}} a^\dagger_{m}b_{m-\hat{\boldsymbol{\delta}}_\alpha}\rb+ h.c.\bigg] \label{genbhz} \nonumber\\&+J_\perp\sum_{m}[a^\dagger_{m}a_{m+\hat{x}(\hat{y})}-b^\dagger_{m}b_{m+\hat{x}(\hat{y})}+h.c.] +\mu_{xy}\sum_{m}[a^\dagger_{m }a_{m}-b^\dagger_{m }b_{m}].
 \end{eqnarray}
    The nearest neighbour (NN) sites are along the diagonals with coordinates  $\hat{\boldsymbol \delta}_1= (\hat{x}+\hat y)/2$ and $\hat{\boldsymbol \delta}_2= (-\hat{x}+\hat y)/2$,  and complex  hopping  parameters,  $\mathcal{J}_l$, where $l=1,2$; $\hat{x}=(1,0)$ and $\hat{y}=(0,1)$; $\Phi$ is a phase factor, which can be regarded as a magnetic flux treading the lattice. The total flux on a square plaquette vanishes just like in Haldane model \cite{fdm}. The next nearest neighbour (NNN) sites are along the horizontal and vertical axes with real  hopping  parameter $J_\perp$. The last term in Eq.~\ref{genbhz} is the staggered onsite potential with  a tuneable parameter  $\mu_{xy}$.   

 Next, we introduce an interlayer coupling  between the porphyrin thin film layers along the $z$-direction. The Hamiltonian  is given by
 \begin{equation}
H_{inter}=J_D\sum_{m} \big[a^\dagger_{m}a_{m+\hat{z}} -b^\dagger_{m}b_{m+\hat{z}}+h.c.\big]  +\mu_z\sum_{m}[a^\dagger_{m }a_{m}-b^\dagger_{m }b_{m}].
\label{genbhz1}
 \end{equation}
 Here, the staggered onsite potential is along the $z$-direction with  tuneable parameter $\mu_z$, and $J_D$ is a real coupling constant.  
 Performing the Fourier transform of the lattice model we obtain $\mathcal{H}=\sum_\bold{k}(a^\dagger_{{\bold k}}, b^\dagger_{{\bold k}})\mathcal{H}(\bold k)(a_{{\bold k}}, b_{{\bold k}})^T$, where
\begin{eqnarray}
&\mathcal{H}(\bold k)=[\rho_1\cos\lb k_+ -{\Phi}\rb+\rho_2\cos\lb k_- +{\Phi}\rb]\sigma_x \nonumber\\&-[\tilde{\rho}_1\cos\lb k_+ -{\Phi}\rb+\tilde{\rho}_2\cos\lb k_- +{\Phi}\rb]\sigma_y\nonumber\\&+[\mu_{xy}-2t_{\perp}\lb\cos ( k_+ + k_-)+\cos  ( k_+- k_-)\rb]\sigma_z\nonumber\\& -\frac{ t_D}{2}[\cos(k_z d)+\cos(k_W d)]\sigma_z; 
\label{fullti}
\end{eqnarray}
The above Hamiltonian Eq.~\ref{fullti} is  obtained with the rescaled parameters $J_\perp\to -t_\perp$, $J_D\to -t_D/4$, and we have fine-tuned the staggered potential to  $\mu_z= J_D\cos(k_W d)$.    We also set the lattice constants $a_x=a_y=1$,  and $a_z=d$, where $d$ is the separation of the  porphyrin thin film layers.  $k_\pm = (k_x\pm k_y)/{2}$, $\rho_1=2\mathscr{R} \mathcal J_1$, $\rho_2=2\mathscr{R}\mathcal J_2$;  $\tilde{\rho}_1=2\mathscr{I}\mathcal  J_1$, and $\tilde{\rho}_2=2\mathscr{I}\mathcal J_2$, where $\mathscr{R}$  and $\mathscr{I}$ denote real and imaginary parts of the complex hopping terms  $\mathcal J_{1,2}$. The model Eq.~\ref{fullti} can be simplified by taking  $\mathcal J_1=\mathcal{J}_2^*$, which implies that $\rho_1=\rho_2=\rho$ and $\tilde{\rho}_1=-\tilde{\rho}_2=\rho$. This is a reasonable simplification and will be adopted throughout our analysis.
\section{2D Weyl semimetal}
As mentioned above, 2D Weyl semi-metals can be constructed  from a lattice model \cite{cas}. In this section, we show how it emerges from the porphyrin thin film layer. In the 2D limit $t_\perp=t_D=0$, the Hamiltonian Eq.~\ref{fullti} has the form
\begin{eqnarray}
\mathcal{H}(k_\pm)&= \rho[\cos\lb k_+ -{\Phi}\rb+\cos\lb k_- +{\Phi}\rb]\sigma_x    \nonumber\\&- \rho[\cos\lb k_+ -{\Phi}\rb-\cos\lb k_- +{\Phi}\rb]\sigma_y. 
\label{2d}
\end{eqnarray}
For $\Phi=0$ or $\pi$, Eq.~\ref{2d} can be written  as
\begin{equation}
\mathcal{H}_{\Phi=0}= \rho\cos\lb\frac{ k_x}{2}\rb\cos\lb\frac{ k_y}{2}\rb\sigma_x  +\rho\sin\lb\frac{ k_x}{2}\rb\sin\lb\frac{ k_y}{2}\rb\sigma_y, 
\label{2dd}
\end{equation}
where $\mathcal{H}_{\Phi=\pi}=-\mathcal{H}_{\Phi=0}$.
\begin{figure}[ht]
\centering
\includegraphics[width=4in]{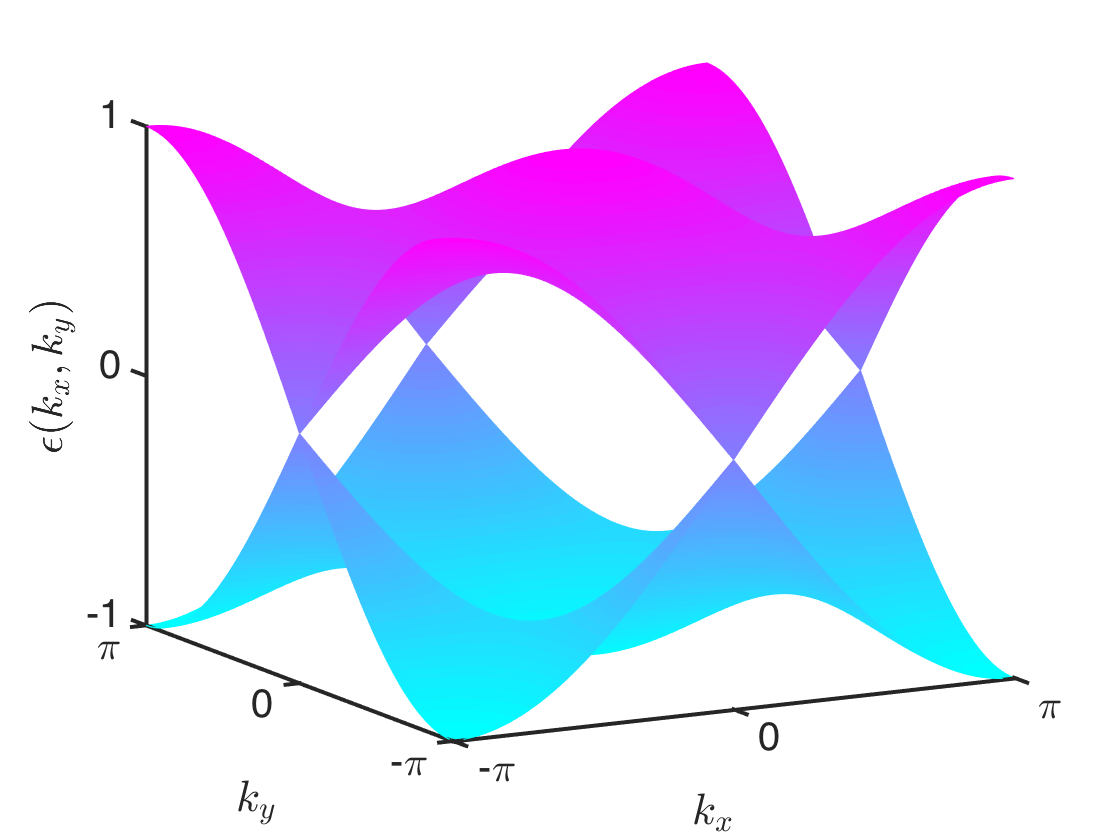}
\caption{Color online. The   energy band of Eq.~\ref{2dd} in units of $\rho$. There are four degenerate points in the BZ with each pair located at $\bold{W}_1=(0,\pm \pi)$ and $\bold{W}_2=(\pm\pi,0)$.}
\label{band1}
\end{figure}
\begin{figure}[ht]
\centering
\includegraphics[width=4in]{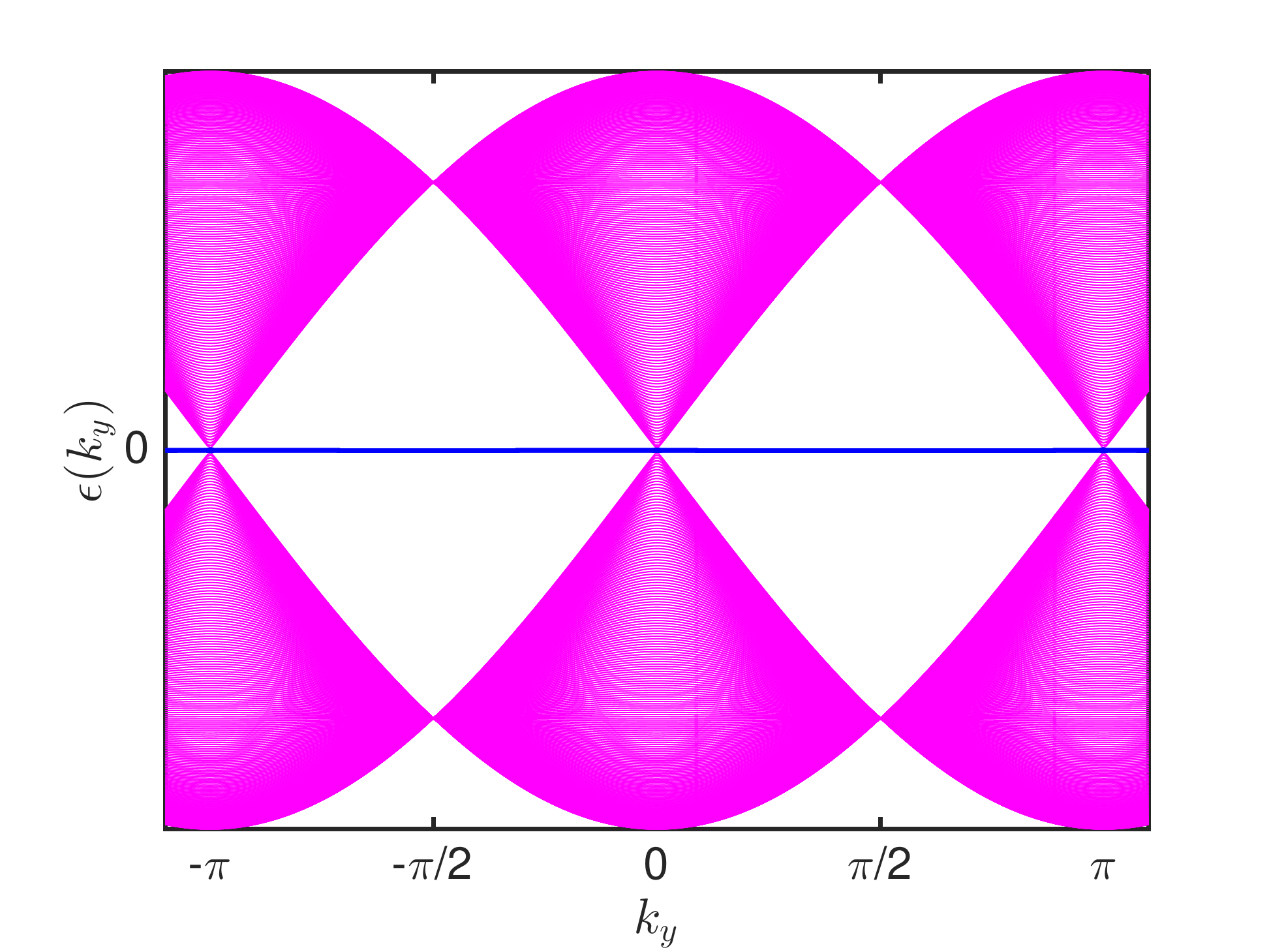}
\caption{Color online. The bulk energy band (pink) and the chiral edge states (blue) of Eq.~\ref{2dd} along the $k_y$ direction. }
\label{2dege}
\end{figure}
\begin{figure}[ht]
\centering
\includegraphics[width=4in]{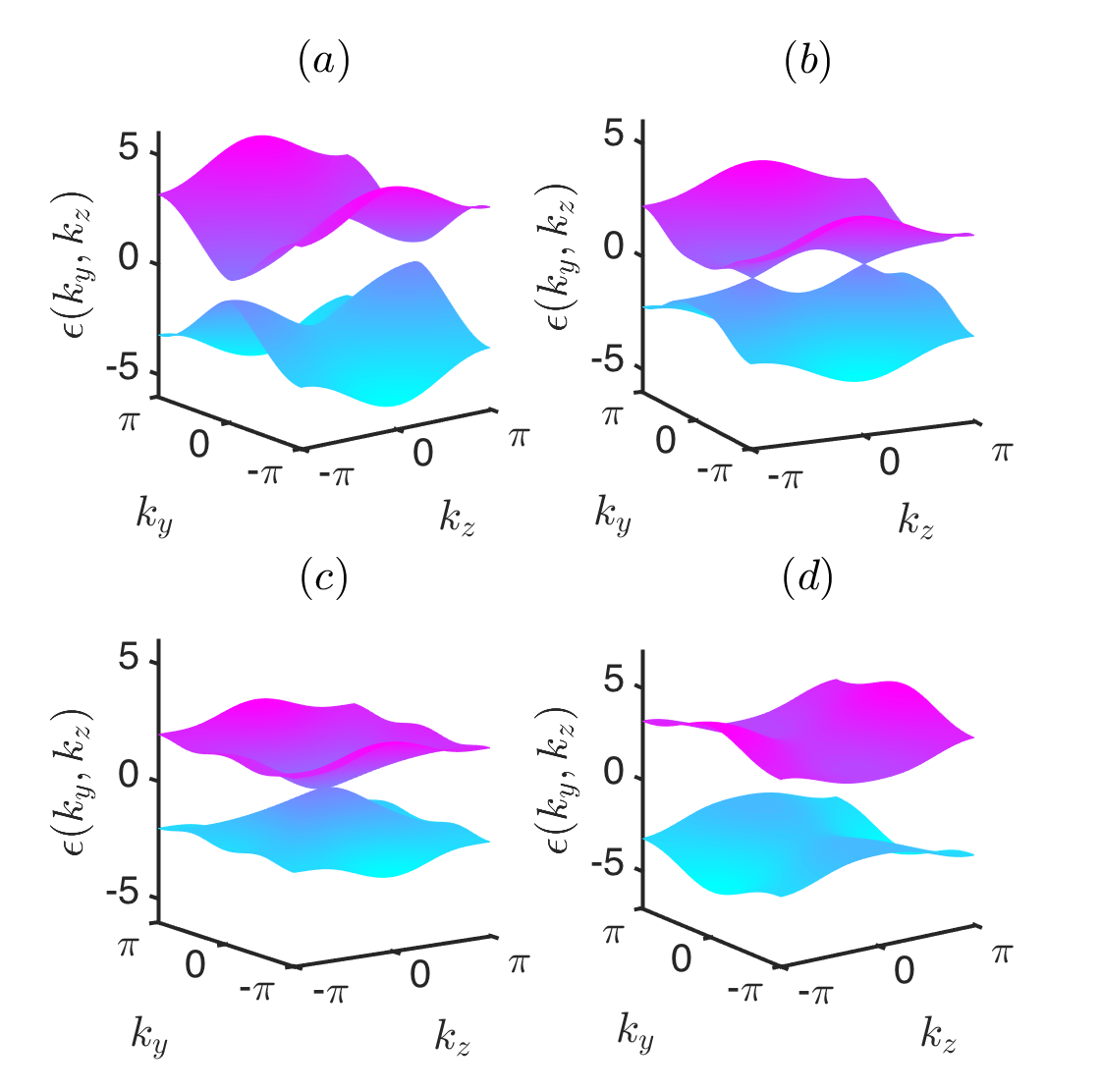}
\caption{Color online. The evolution of the energy along $k_x=0$ at $\Phi=\pi/2$ with $\mu_{xy}=4t_\perp$. The four regimes are: the insulating phase $(a). ~\gamma<\gamma_-$, Weyl semi-metallic phase $(b).~ \gamma_-<\gamma<\gamma_+$, phase transition point $(c).~\gamma=\gamma_+$, and the 3D QAH phase $(d).~\gamma>\gamma_+$. }
\label{band}
\end{figure}
\begin{figure}[ht]
\centering
\includegraphics[width=4in]{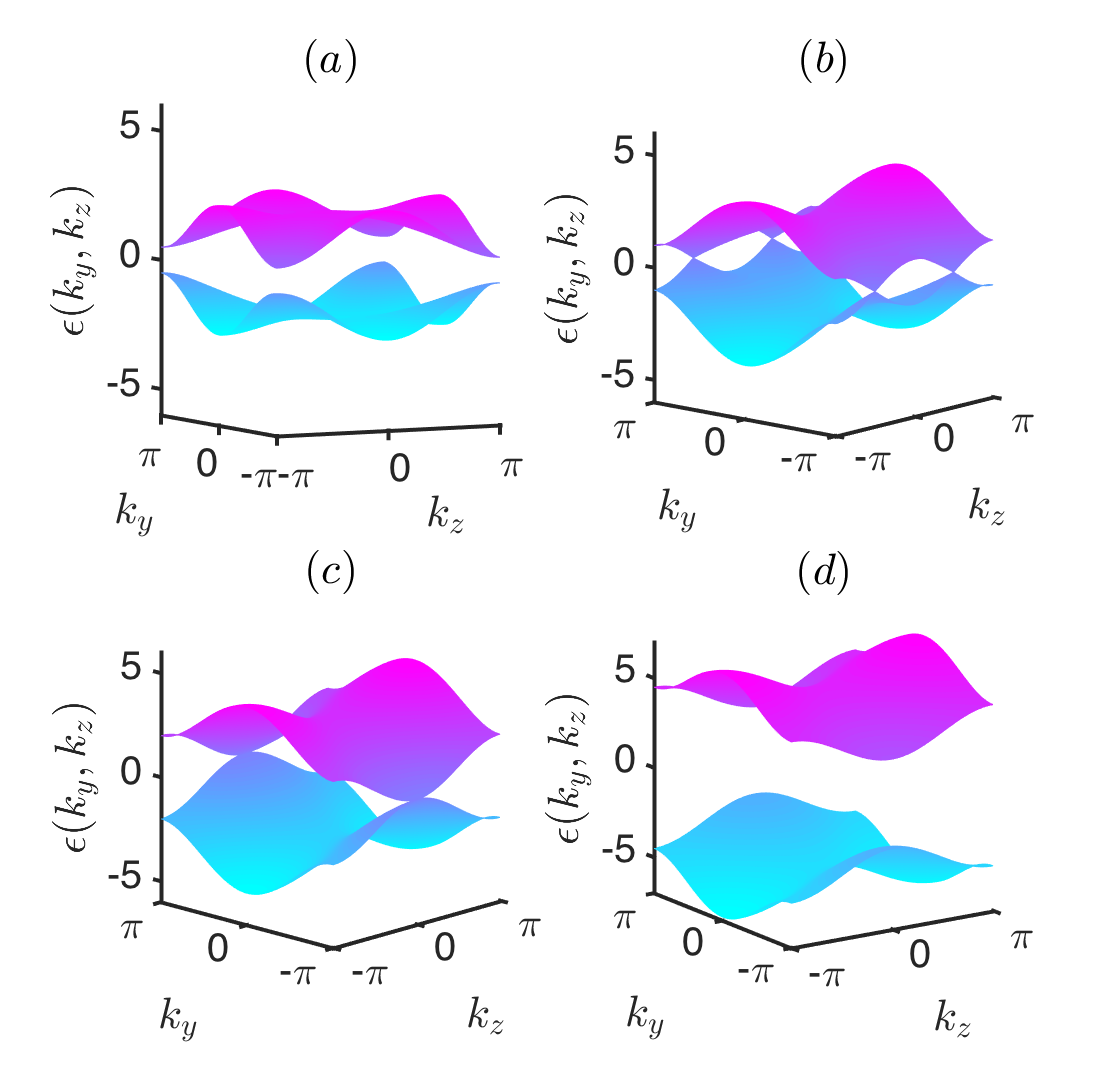}
\caption{Color online. The evolution of the energy along $k_x=0$ at $\Phi=0$ or $\Phi=\pi$ with $\mu_{xy}=0$. The parameters are the same as Fig.~\ref{band}.}
\label{band2}
\end{figure}
 \begin{figure*}[ht]
\centering
\includegraphics[width=6in]{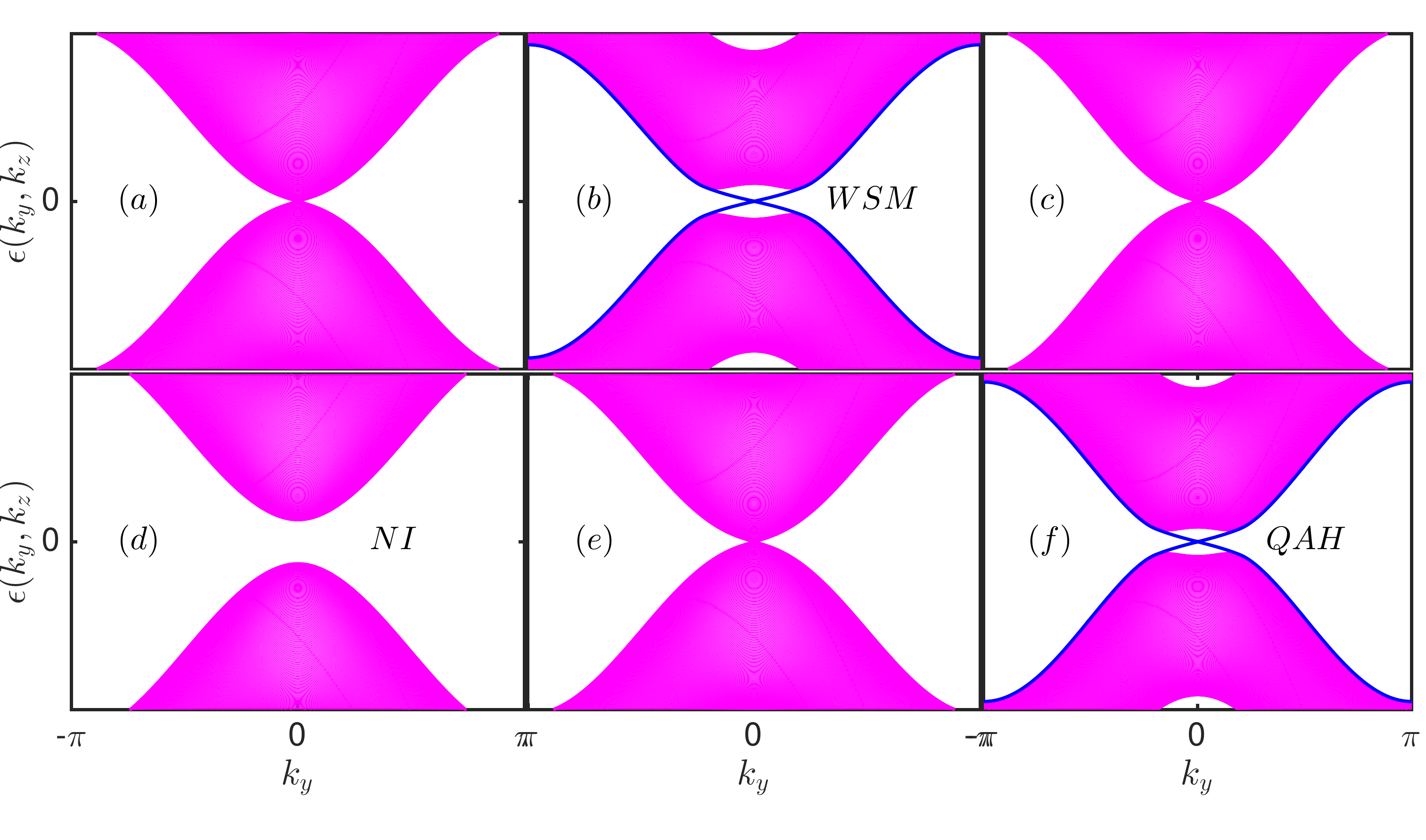}
\caption{Color online.  The bulk energy band and the surface states of Eq.~\ref{fullti} along the $k_y$ direction with several values of $k_z$ with $\Phi=\pi/2$. From $(a)-~(c)$, describe a Weyl semi-metallic phase with fine-tuned $\mu_{xy}=4t_\perp$,   $t_D=0.75t_S$, $\gamma_-<\gamma=0.25t_S<\gamma_+$, $t_\perp=t_S=1$;  and $k_z$ varies as $k_z=k_z^-$, $k_z=k_z^-/2$, and $k_z=-k_z^-$ or $k_z^+$ respectively.  From $(d)-~(f)$, describe the insulating phases, now $t_\perp<0$ see Eqs.~\ref{QAH} and \ref{NI}. (d) $\gamma<\gamma_-$ and $k_z=\pi$. (f) $\gamma=\gamma_+$ and $k_z=0$. (e) $\gamma>\gamma_+$ and $k_z=0$. See text for explanation.}
\label{Full_TI_edge}
\end{figure*}
  As shown in Fig.~\ref{band1}, the energy band  has four degenerate points  located at $\bold{W}_1=(0,\pm \pi)$ and $\bold{W}_2=(\pm\pi,0)$. However, the degeneracy of an energy band does not guarantee a Weyl semi-metallic phase. To obtain a nontrivial topological semimetal,  symmetry consideration must be taken into account. For the Hamiltonian in Eq.~\ref{2dd},  time-reversal symmetry ($\bold k\to -\bold k;~\boldsymbol{\sigma}\to-\boldsymbol{\sigma}$) is broken but inversion symmetry ($\bold k\to -\bold k$) is preserved.
   For  2D systems, however, there is an additional hidden discrete symmetry with an anti-unitary operator \cite{cas}. More generally,  {\textit if a system is invariant under the action of an anti-unitary operator and the square of the operator is not equal to 1, there must be degeneracy protected by this anti-unitary operator} \cite{cas}. In the present model, there is an anti-unitary operator for which the Hamiltonian (Eq.~\ref{2dd}) is invariant. It is  given by $\mathcal{U}=\sigma_x \mathcal{K} T_{\hat x/2+ \hat y/2}$, where $\mathcal{K}$ is complex conjugation and $T_{\hat x/2+ \hat y/2}$ translates the lattice by $\hat x/2$ and $\hat y/2$  along the $x$- and $y$-directions. It is easy to check that $\mathcal{U}^{-1}\mathcal{H}_{\Phi=0} \mathcal{U}=\mathcal{H}_{\Phi=0}$. It follows that  $\mathcal{U}^2=e^{-i (k_x+k_y)}=-1$ at $\bold{W}_{1}$ and  $\bold{W}_{2}$. Note that $\mathcal{U}^2=-1$ at various points in the BZ, e.g., $\bold k=(\pi/2,\pi/2)$. However, the energy does not vanish at these points, the reason being that they are not $\mathcal{U}$-invariant points. Thus, the theorem stated above is only valid at the $\mathcal{U}$-invariant points.  The four degenerate points in the energy spectrum is consistent with  Nielsen-Ninomiya theorem \cite{niel}, which states that  Weyl points must occur in pair(s) with opposite helicity in a lattice model.  Near these points, the Hamiltonian is linearized as
\begin{equation}
\mathcal{H}_1(\bold{q})=v_F[\mp q_y\sigma_x\pm q_x\sigma_y],\label{2dn}~
\mathcal{H}_2(\bold{q})=v_F[\mp q_x\sigma_x\pm q_y\sigma_y],
\end{equation}
where $\bold{q}=\bold{k}-\bold{W}_{1,2}$ and $v_F=\rho/2$. 

The Hamiltonian has the general form $\mathcal{H}(\bold{q})=\sum_{ij}v_{ij}q_{i}\sigma_j$, where $v_{ij}$ form a  $2\times 2$ matrix.  The  chirality of the Weyl points is given by 
\begin{equation} \chi_\pm=\textrm{sign}\lb \textrm{det}[v_{ij}]\rb\label{chi}.\end{equation} 
From Eqs.~\ref{2dn} and \ref{chi} we obtain $\chi_\pm=\pm 1$ for the whole system, which signifies the topological nature of the system. As a massless Dirac fermion with chirality, the system above can be regarded as a 2D Weyl semi-metal which hosts a 2D Wely fermion.  In Fig.~\ref{band1}, opposite chirality is assigned to neighbouring Weyl points in cyclic order.  Moreover, in 2D Weyl semimetal there is a chiral edge state propagating in the intermediate region between the  Weyl points. This can be explicitly shown by considering a semi-infinite system  with periodic boundary conditions along the $k_y$ direction and open boundary  condition along the $k_x$ direction  \cite{zhang}.  The  bulk band  is shown in Fig.~\ref{2dege}  along the $k_y$ direction. The bulk gap vanishes at the locations of the Weyl points along the $k_y$-axis consistent with Fig.~\ref{band1}. However, topological protected flat-band chiral edge states  emerge in-between the Weyl nodes. These chiral edge states connect the Weyl points with opposite chirality along the $k_y$-direction.

\section{3D Weyl semimetal}
In this section, we study the possibility of 3D Weyl semi-metallic phase in the proposed lattice model. The goal is to utilize this model to simulate the TI multilayer surface states (Fermi arc). In 3 dimensions we must have $t_D\neq0$; a 3D Weyl semimetal can be obtained with a judicious  choice of $\Phi$. In particular,   for $\Phi=\pi/2$ and $\mathcal J_1=\mathcal{J}_2^*$,   Eq.~\ref{fullti} has the form $\mathcal{H}=\mathcal{H}_{\Phi=\pi/2} +{\mathcal{H}_z}$, where
\begin{equation}
\mathcal{H}_{\Phi=\pi/2}= 2v_F\cos\lb\frac{ k_x}{2}\rb\sin\lb\frac{ k_y}{2}\rb\sigma_x-2v_F\cos\lb\frac{ k_y}{2}\rb\sin\lb\frac{ k_x}{2}\rb\sigma_y.\label{3dd}
\end{equation}
\begin{equation}
{\mathcal{H}_z}=\Big[\mu_{xy}-2t_{\perp}(\cos k_x+\cos k_y)-\frac{ t_D}{2}\lb\cos(k_z d)+\cos(k_W d)\rb\Big]\sigma_z.
\label{3d}
\end{equation}
 It is easy to see that with  a fine-tuned $\mu_{xy}=4t_\perp$, the partial continuum limit of Eqs.~\ref{3dd} and \ref{3d}  is exactly the inner $2\times 2$ block of Eq.~\ref{par} and the Weyl nodes are located at the same points $\bold W=(0,0,k^\pm_z)$, with $k_z^\pm=\pi/d \pm k_W$. Thus, the porphyrin thin film multilayer lattice model recovers that of TI thin film multilayer in the partial continuum limit.  The evolution of the energy bands in the BZ are  shown in Fig.~\ref{band}. Near the Weyl points the Hamiltonian is given by
 \begin{equation}
 \mathcal H(\bold q) = v_Fq_y\sigma_x-v_Fq_x\sigma_y\mp \tilde{v}_F q_z\sigma_z,
 \label{3dn}
 \end{equation}
where $\bold q= \bold k-\bold W$, and $\tilde{v}_F=t_D d\sin(k_Wd)/2$.

 The Hamiltonian still has the general form $\mathcal{H}(\bold{q})=\sum_{ij}v_{ij}q_{i}\sigma_j$, only that $v_{ij}$ is now a $3\times 3$ matrix with components $v_{yx}=v_F,~v_{xy}=-v_F,~v_{zz}=\pm\tilde{v}_F$. The chirality of the Weyl points is the same $\chi_\pm =\pm 1$.  In this case,  the nontrivial topology of Eq.~\ref{fullti} stems from the fact that Eq.~\ref{fullti} preserves inversion symmetry but breaks time-reversal symmetry, when $\Phi=\pi/2$. Another judicious choice of $\Phi$ is $\Phi=0$ or $\pi$.  The resulting Hamiltonian in this case is given by $\mathcal{H}=\mathcal{H}_{\Phi=0} +{\mathcal{H}_z}$, but it is different from that of TI thin film. However, the system still preserves inversion symmetry and breaks time-reversal symmetry; thus a Weyl semi-metallic phase can be obtained.   There are four Weyl points in the BZ, each pair is located at $\bold W_1=(0,\pi,\tilde{k}^\pm_z)$ and $\bold W_2=(0,-\pi,\tilde{k}^\pm_z)$, where $\tilde{k}^\pm_z=\pi/d \pm \tilde{k}_W$ and $
 \tilde{k}_W= \frac{1}{d}\arccos[\cos(k_Wd)-\tilde{\mu}_{x,y}]$, with $\tilde{\mu}_{x,y}=2\mu_{x,y}/t_D$. The energy bands are shown  in Fig.~\ref{band2}. The Hamiltonian near the Weyl points  is a combination of Eq.~\ref{2dn} and the last term in  Eq.~\ref{3dn}.

Now, we study the surface states evolution of the Weyl semi-metallic phases above. This is an important feature of 3D Weyl semimetals \cite{wan, kre} and it is what is observed in most experiments  \cite{llu, Xu,lv, lv1}. In our lattice model,   these states can be solved explicitly  for any surface not perpendicular to the $z$-axis.  In fact, they are nothing but  the edge states of the effective 2D model for fixed values of $k_z$.  We have shown the evolution of the states for $\Phi=\pi/2$  in Fig.~\ref{Full_TI_edge} (a)--(f), which corresponds exactly to the ultra-thin film of TI multilayer studied above.  The top panel describes the Weyl semi-metallic phase bounded by two gapless bulk bands at the location of the Weyl points. For $k_z\in ( {k}^-_z,  k^+_z)$, there exist  dispersive surface states propagating in the vicinity of the bulk gap only when $t_\perp>0$. They are gapless at $k_y=0$ exactly at zero energy. In the bottom panel we show the insulating phases after the Weyl nodes annihilate and a gap opens at $k_z=0$ or $k_z=\pi/d$. In this case,  the surface states still capture the appearance of the two insulating phases -- 3D QAH and NI only when $t_\perp<0$. These results are  consistent with our previous analysis and the energy dispersion in Fig.~\ref{band}. For other choices of $\Phi$ such as $\Phi=0,~\pi$,  the situation is a little bit different. The gapless surface states only occur at $k_y=\pm\pi$, when $k_z\in (\tilde{k}^-_z, \tilde{k}^+_z)$, but $k_y=0$ is gapped in this case and we observe that there exist gapped surface states propagating in this vicinity (not shown).

\section { Conclusion}
In this paper, we have presented a detail analysis of two thin film models of Weyl semimetals. We showed that in an ultra-thin film of topological insulator multilayer the    parameters of the system can change sign as the system transits from one topological phase to another. In this model, we presented the low-temperature dependence of the chiral magnetic conductivity, induced by a time-dependent magnetic field. We showed that the topological phases of the system can, indeed, be captured  by the plateaus of the chiral magnetic conductivity.  We also proposed and studied a simple lattice model of porphyrin thin film.  We showed that this model embodies many Weyl semi-metallic phases for a specific gauge choice, which acts as a magnetic flux treading the lattice. We obtained a 2D Weyl semi-metallic phase in the $\sigma_x$-$\sigma_y$ space. We showed that the degeneracy of the Weyl nodes is protected by an anti-unitary operator.   Our model also realized a 3D Weyl semi-metallic phase, which can be regarded as the lattice model for an ultra-thin film of topological insulator (TI) multilayer.  Thus, it paved the way to numerically study the surface states of the TI multilayer.  We obtained the edge states and the surface states in two  and three dimensions respectively, as well as in all the nontrivial topological phases of the TI multilayer in three dimensions.  As the  porphyrin thin film is an organic material that can be grown in the laboratory, the proposed model can perhaps be studied experimentally or in 2D optical lattices. As shown in this paper, the porphyrin thin film is also a candidate to search for chiral relativistic fermions in two dimensions.    

\section*{ Acknowledgement}
 The author would like to thank J. -M. Hou for enlightening discussions. The author would also like to thank African Institute for Mathematical Sciences for hospitality. Research at Perimeter Institute is supported by the Government of Canada through Industry Canada and by the Province of Ontario through the Ministry of Research
and Innovation.

\end{document}